\documentstyle[12pt]{article}

\begin{document}

\hoffset = -1truecm
\voffset = -2truecm
\baselineskip = 10 mm

\title{\bf A unity of QCD evolution dynamics at small $x$ range}

\author{
{\bf Wei Zhu}, {\bf Zhenqi Shen}, {\bf Jifeng Yang} and {\bf Jianhong Ruan} \\
\normalsize Department of Physics, East China Normal University,
Shanghai 200062, P.R. China \\
}

\date{}

\newpage

\maketitle

\vskip 3truecm

\begin{abstract}

     The DGLAP, BFKL, modified DGLAP and modified BFKL equations
are constructed in a unified partonic framework.  The
antishadowing effect in the recombination process is emphasized,
which leads to two different small $x$ behaviors of gluon
distribution. In the meantime, the BFKL equation and modified
DGLAP equation are viewed as the corrections of the initial gluon
correlations to the evolution dynamics at twist-2 and twist-4,
respectively. A partonic explanation of Regge theory for the BFKL
dynamics and the relation of the BFKL dynamics with the helicity
configuration of the initial gluons are presented.

\end{abstract}


$keywords$:  antishadowing; evolution equations; QCD recombination
process.

\newpage
\begin{center}
\section{Introduction}
\end{center}

    The evolution equations for parton distributions play an important
role in the QCD studies of high-energy processes. A standard
evolution equation is the DGLAP equation (by Dokshitzer, Gribov,
Lipatov, Altarelli and Parisi in [1]), which sums large
logarithmic $Q^2$ corrections ($LL(Q^2)$) to the integrated parton
distributions. The DGLAP equation can be obtained by using the
renormalization group or graphic techniques. The DGLAP equation
successfully predicts the $Q^2$-dependence of parton distributions
in a broad kinematic range. However, the DGLAP equation is
unsatisfactory at small $x$, where the resummation of the leading
logarithm of $x$ ($LL(1/x)$) becomes important and it leads to the
BFKL equation (by Balitsky, Fadin, Kuraev and Lipatov in [2]). The
BFKL equation for the unintegrated gluon distribution is
originally derived by using the Reggeization of gluons.  Both the
BFKL and DGLAP equations predict a rapid increase of the parton
densities at small $x$ due to parton splitting, and the unitarity
limit is violated. Therefore, the corrections of the higher order
QCD, which suppress or shadow the growth of parton densities,
become a focus of intensive study in recent years. Several
nonlinear evolution equations are proposed at small $x$. Some of
them are the GLR-MQ equation  (by Gribov, Levin and Ryskin in [3]
and by Mueller and Qiu in [4]), modified DGLAP equation (by Zhu
and Ruan in [5]), JIMWLK equation (by Jalilian-Marian, Iancu,
McLerran, Weigert, Leonidov and Kovner in [6]), Balitsky-Kovchegov
equation [7], and various versions of the evolution equations
based on the color dipole picture [8]. Unfortunately, we still
lack a unified picture to clearly understand the relation between
the DGLAP and BFKL equations and the unitarity corrections on
them.

     In this paper we try to present a unified partonic framework for understanding
some of QCD evolution equations at small $x$. Our idea is
straightforward.  We begin with an elementary amplitude Fig. 1a of
deep inelastic scattering (DIS) in the parton scattering picture,
where the dashed line implies a virtual current probing gluon.
This amplitude, together with its conjugate amplitude, constructs
the DGLAP equation for gluons. The correlations among initial
partons are neglected in the derivation of the DGLAP equation.
Obviously, this assumption is invalid in the higher density region
of partons, where the parton wave functions begin to spatially
overlap. Therefore, the leading corrections of the correlating
initial gluons to the elementary amplitude at small $x$ should be
considered. To this end, we add the possible initial gluons to
Fig. 1a step by step.  The resulting three sets of amplitudes are
listed in Fig. 1b-1d.  It is interesting that these amplitudes
produce the BFKL equation, modified DGLAP equation and modified
BFKL equation, respectively.

      A key problem in the restoring unitarity is the origin of the negative
corrections. In the viewpoint of elementary QCD interaction, the
suppression to the gluon splitting is naively from its inverse
process-the gluon fusion (or the recombination process, i.e., a
gluon splitting combines with the gluon fusion). DIS structure
functions are the imaginary parts of the amplitudes for the
forward Compton scattering of the target with a probe.  According
to the time-ordered perturbative expansion  of the statement of
unitarity of the S-matrix, the structure functions are associated
with the sum of cut diagrams. These different cut graphs represent
various possible sub-partonic processes due to the unitarity of
the perturbative S-matrix. Therefore, the sum of cut graphs in the
recombination process is necessary not only for infrared safety,
but also for restoring unitarity. As Gribov, Levin and Ryskin
pointed out that the $2\rightarrow 2$--$2\rightarrow 2$
cut-diagram of the recombination process always contributes the
positive correction and it can not lead to the negative effect in
parton evolution equation [3].  The negative screening effect in
the recombination process can occur in the interferant or virtual
cut-diagrams of the recombination amplitudes [5]. For computing
the contributions from the interference processes, work [3] used
the AGK cutting rule [9]. The result is the GLR-MQ equation. The
GLR-MQ equation was generalized to include the multi-gluon
recombination process in the Glauber-Mueller formula [8].

      In the next advance, one of us (zhu) disputed in [5] the above
mentioned applications of the AGK cutting rule in the GLR-MQ
equation. The TOPT-cutting rule based on the time ordered
perturbation theory (TOPT) replaces the AGK cutting rule to expose
the relations among various cut diagrams in a recombination
process [5]. Thus, we can completely compute the contributions of
the gluon recombination process. Following it, the modified DGLAP
equation was proposed in [5]. A remarkable property of this
equation is that the positive antishadowing and negative shadowing
in the nonlinear evolution equation are naturally separated. The
recombination among the gluon at Bjorken variable $y\ge x$ will
reduce the density of the gluon at $x$. In the meantime, a gluon
in the region $[x/2,x]$ can generate a gluon at $x$ if the former
combines with its companion. In consequence, the modifications in
the gluon distribution at $x$ could come from either the screening
(shadowing) effect of the fusion of two gluons at $y\ge x$, or the
enhancement (antishadowing) effect of the fusion of two gluons at
$y\ge x/2$ and $y<x$. As a result, the corrections arising from
gluon recombination at small $x$ depends not only on the size of
gluon density at this value of $x$, but also on the sharp of gluon
density in the region $[x/2,x]$. Thus, the shadowing effect in the
evolution process will be weakened by the antishadowing effect if
the gluon distribution has a steeper form [5,10]. Although the
gluon recombination re-distributes the parton densities in the
above mentioned processes, it always keeps the momentum
conservation. We note that the GLR-MQ equation also computes the
positive contribution of the $2\rightarrow2$-$2\rightarrow2$ cut
diagram, the momentum conservation of the recombination process is
violated due to the mistakes in the application of the AGK cutting
rule, where the contributions of all cut diagrams only differ by a
numeric weight.

     We shall extend this work along the above mentioned
researches. For this purpose, we give a short review about the
applications of the TOPT in the factorization scheme. Following it
the QCD evolution equations listed in Fig. 1 are derived in a
traditional Bjorken frame. We find that the antishadowing effect,
which is a direct result of the momentum conservation, is
non-negligible in the corrections of the gluon recombination to
the DGLAP and BFKL equations. In particular, we predict that the
gluon distribution asymptotically (abruptly) approaches to the
saturation limit below (above) a moderate $Q^2$-scale due to the
antishadowing effect.

    It is interesting that, in this framework, the BFKL equation can be
derived as a correction of the correlations of initial gluons at
the twist-2, and the relation of the BFKL dynamics with the
helicity configuration of the initial gluons becomes transparent.
Then the partonic interpretation of Regge theory for the BFKL
dynamics is presented.  We also show that the similarity between
the modified DGLAP equation and the BFKL equation in their ladder
diagrams.  We compare the modified BFKL equation with the color
dipole model of the Balitsky-Kovchegov equation.  We find that
they have a similar physical picture but they are really different
evolution dynamics. As is well known, the QCD evolution dynamics
for parton densities is extensively applied in high energy
physics. The discussions in this work present a new way to reveal
some unknown properties, which can be regarded as a supplement to
our knowledge about evolution equations.

      The paper is organized as follows.  In Section 2 we discuss
the separation of the evolution kernels from the TOPT diagrams
using the semi-classical Weizs$\ddot{\rm a}$cker-Williams ($W-W$)
approximation [11].  In Sections 3 we discuss the connection
between the DGLAP equation and BFKL equation.  In Section 4 the
modified DGLAP equation is generalized to include the BFKL
equation. The discussions and concluding remarks are given in
Section 5.

\newpage
\begin{center}
\section{Factorization}
\end{center}

    The partonic picture of a high energy process is frame- and
gauge-dependent.  The appropriate choices of the coordinate frame
and gauge are helpful for extracting the evolution kernels using a
simple partonic picture.  A partonic scattering model takes
Bjorken frame, i.e., the momentum of the nucleon is

$$P^{\mu}=(P_0, \underline{P}, P_3)=(P,\underline{0},P),\eqno(2.1)$$
while the virtual probe has almost zero-energy $q_0=
(2M\nu+Q^2)/4P$ and zero-longitudinal momentum
$q_3=-(2M\nu-Q^2)/4P$ as $P\rightarrow \infty$, so that the
momentum of the probe is mainly transverse to the nucleon
direction.  In the mean time, we work in the physical axial gauge:
$n\cdot A=0$, with $n$ being the light cone vector

$$n^{\mu}\equiv \frac{1}{\sqrt{2}}(1,\underline{0},-1),\eqno(2.2)$$
and

$$\overline{n}^{\mu}\equiv \frac{1}{\sqrt{2}}(1,\underline{0},1). \eqno(2.3) $$

    The evolution kernel in a QCD evolution equation is a part of more
complicated scattering diagram. In general, the correlations among
the initial partons make these partons off mass-shell. One of the
most important hypotheses in the derivation of the DGLAP equation
is that all correlations among initial partons are negligible
during the interaction.  In this case, the interaction of a
virtual probe with the nucleon can be factorized as the
nonperturbative parton distribution and hard probe-parton matrix
in the collinear factorization scheme.

    On the other hand, at the higher density range of partons, these
correlations among the initial partons can no longer be neglected
(Fig. 2). In this case, the transverse momenta of the initial
partons are non-zero and these partons are generally off
mass-shell, therefore, the $k_T$-factorization scheme [12] is
necessary.

     However, in this work we use the $W-W$ approximation rather than
the $k_T$-factorization scheme.  The reason is that the $W-W$
approximation allows us to extract the evolution kernels and to
keep all initial and final partons of the evolution kernels on
their mass-shell.  For illustrating this idea, we consider a
sub-process as shown in Fig. 3 , where the momenta of three
massless particles are parameterized as

$$p_1=(p,\underline{0},p),$$

$$p_2=(x_2p+\frac{\underline{k}^2}{2x_2p}, \underline{k}, x_2p), $$
and

$$p_3=(p_3^0, -\underline{k}, x_3p), \eqno(2.4) $$
where $p\rightarrow \infty$. The particle 3 is off its mass-shell.
We decompose this propagator into a forward and a backward
propagator according to the TOPT. The corresponding momenta are on
mass-shell

$$\hat{p}_F=(x_3p+\frac{\underline{k}^2}{2x_3p}, -\underline{k}, x_3p),$$
and

$$\hat{p}_B=(-x_3p-\frac{\underline{k}^2}{2x_3p}, -\underline{k}, x_3p), \eqno(2.5)$$
respectively. Note that Eqs. (2.4) and (2.5) require

$$xp\gg \mid \underline{k}\mid, \eqno(2.6)$$
which is consistent with the conditions of high energy- and small
angle-scattering in the $W-W$ approximation. One can find that the
forward propagator contributes the power-factor $1/k^2_{\perp}$,
since

$$\frac{1}{2E_3}\frac{1}{E_2+E_{\hat{p}_F}-E_1}\sim\frac{1}{\underline{k}^2},\eqno(2.7)$$
while the backward propagator gives the contributions $\sim
1/p^2$. If the dominant contributions in a process are from the
terms with the power of transverse momentum, one can find that the
contributions from the backward propagator are suppressed at
$p\rightarrow \infty $, and the dominant contribution is from the
forward component. In this case, we can break the forward
propagator since it propagates on mass-shell.

       The wave function of a parton in the Bjorken frame occupies a space with
transverse size $\sim 1/\underline{k}$ and longitudinal size $\sim
1/xp$ in the nucleon. The partons are regarded as independent if
the partons are dilute and the parton distributions obey the DGLAP
equation (Fig. 4a). However, this evolution equation predicts a
rapid rise of gluon multiplicities inside the nucleon at small
$x$. At sufficiently large gluon densities, the wave functions of
the gluons overlap with each other.  We call such a correlating
gluon cluster with a transverse scale $\underline{x}_{ab}$ as the
cold spot, which phenomenologically describes the correlation
among initial partons (the dashed circles in Fig. 4b). In this
work we mainly consider the cold spots consisted of two gluons,
which dominates the leading small $x$ processes.

    The elementary correlation among two initial partons in a cold spot
is shown in Fig. 5, where the dark circle implies all possible
QCD-channels. Let us consider a configuration of two correlating
partons that form a cold spot inside a nucleon. We assume that the
longitudinal momentum of the initial parton is much larger than
its transverse momentum in the Bjorken frame even at small $x$.
Thus we can take the $W-W$ approximation and extract the two
parton correlation function from the TOPT-diagram as

$$f(\underline{p}_a, \underline{p}_b,x_a,x_b)$$
$$\equiv \frac{E_{p_a}+E_{p_b}}{2E_P}\vert M_{P\rightarrow
p_ap_bX}\vert^2
\left[\frac{1}{E_P-E_{p_a}-E_{p_b}}\right]^2\left[\frac{1}{2E_{p_a}}\right]^2
\left[\frac{1}{2E_{p_b}}\right]^2\prod_X\frac{d^3k_X}{(2\pi)^32E_X}.
\eqno(2.8)$$

    Using Fourier transform with respect to the transverse momentum
transfer, these partons localize in a range with the size
$\underline{x}_{ab}$ on the impact space.  The unintegrated
distribution $f(x_{ab}^2;x_a,x_b)$ can be understood as the
probability of finding two partons with transverse distance
$\underline{x}_{ab}$ and the longitudinal momentum  fractions
$x_a$ and $x_b$, respectively.

     The probability of forming a cold spot is proportional to
the "size'' of gluon. In particular, we have a straightforward
bound condition

$$f(\underline{x}_{ab}^2=0,x_a,x_b)=0, \eqno(2.9)$$
which implies that the overlap probability of two size$less$
partons in the impact space vanishes.

    In the impact space, the integrated gluon distribution reads

$$G(\underline{x}_{Q^2}^2,x)\equiv xg(\underline{x}_{Q^2}^2,x)=\int_{\underline{x}_{Q^2}^2}^{R^2_N}
\frac{d\underline{x}_{ab}^2}{\underline{x}_{ab}^2}f(\underline{x}_{ab}^2,x),\eqno(2.10)$$
or

$$f(\underline{x}_{ab}^2,x)=-\underline{x}_{Q^2}^2\frac{\partial xg(\underline{x}_{Q^2}^2,x)}{\partial \underline{x}_{Q^2}^2}
\left|_{\underline{x}_{Q^2}^2=\underline{x}_{ab}^2}\right.,
\eqno(2.11)$$ while in the momentum space, they are

$$G(Q^2,x)\equiv xg(Q^2,x)=\int^{Q^2}_{Q^2_{min}}\frac{d\underline{k}^2}{\underline{k}^2}
f(\underline{k}^2, x), \eqno(2.12)$$ or

$$f(\underline{k}^2, x)=Q^2\frac{\partial xg(Q^2,x)}{\partial
Q^2}\left |_{Q^2=\underline{k}^2}\right ., \eqno(2.13)$$ where
$R_N$ is an effective radius of the nucleon.  Note that the
momentum $Q$ of the probe is mainly transverse to the nucleon
direction.

        One can imagine that the cold spots fill the whole transverse
plane of a nucleon at a higher density scale $n_s$, which
corresponds to the saturation of the gluon distributions (Fig.
4c). The saturation is a complicated state, where the long- and
short-distance correlations among partons coexist in a nucleon. In
this work, we focus on the region far from the saturation limit.

    In the next step, we discuss the separation of the virtual
probe-vertex from the DIS amplitude. In principle, in- and
out-probe lines can be attached to the left- and right-hand sides
of the cut line in all possible ways. However, in the Bjorken
frame, one only need to pick up the diagrams with two photon lines
following the cut line at the leading logarithmic ($Q^2$)
approximation ($LL(Q^2)A$), since these diagrams contribute the
dominant terms $\sim d\underline{k}^2/\underline{k}^2$ [5].

    Under the same (i.e., $LL(Q^2)A$) condition, the
contributions of the backward components in the two propagators
connecting with the probe are suppressed. Thus, we can take the
$W-W$ approximation to factorize the virtual probe-vertex from the
cut diagram and extract the evolution kernel in TOPT. However, the
BFKL dynamics works in the region beyond the $LL(Q^2)A$.
Fortunately, we could show that the dominant contributions still
come from the terms with the power of the transverse momentum,
i.e., we pick up the dominant terms $\sim
d\underline{k}^2/\underline{k}^n$. Therefore, the $W-W$
approximation is still feasible in the separation of the
probe-vertex for the BFKL dynamics.

\newpage
\begin{center}
\section{Evolution equations at twist-2}
\end{center}

    Using the $W-W$ approximation mentioned in Sec. 2, the
DIS cross section at leading twist is generally factorized into
the following convolution form

$$d\sigma(probe^* p\rightarrow k'X)$$
$$=f(\underline{k}^2_1,x_1)\otimes{\cal
K}\left(\frac{\underline{k}^2_2}{\underline{k}_1^2},\frac{x_2}{x_1},\overline{\alpha}_s\right)\otimes
d\sigma(probe^*k_2\rightarrow k')$$
$$\equiv \Delta f(\underline{k}^2_2,x_2)\otimes d\sigma(probe^*k_2\rightarrow
k'), \eqno(3.1)$$  with the perturbative evolution kernel ${\cal
K}$, the $probe^*$-parton cross section
$d\sigma(probe^*k_2\rightarrow k')$ and the nonperturbative
unintegrated parton distribution function(s) $f$.  For simplicity,
we take the fixed $\alpha_s$ in this work. We emphasize that the
factorization between ${\cal K}$ and
$d\sigma(probe^*k_2\rightarrow k')$ is valid because the
propagators linking these two parts are on mass-shall in the $W-W$
approximation, and hence we can define $\Delta
f(\underline{k}^2_2;x_2)$. According to the scale-invariant parton
picture of the renormalization group [13], we regard $\Delta
f(\underline{k}^2_2,x_2)$ as the increment of the distribution
$f(\underline{k}^2_1,x_1)$ when it evolves from $(\underline
{k}^2_1, x_1)$ to $(\underline {k}^2_2, x_2)$. Thus, the
connection between the unintegrated gluon distribution functions
$f(\underline{k}^2_1,x_1)$ and $f(\underline{k}^2_2,x_2)$ due to
Eq. (3.1) is

$$f(\underline{k}^2_2,x_2)=f(\underline{k}^2_1,x_1)+\Delta
f(\underline{k}^2_2,x_2)$$
$$=f(\underline{k}^2_1,x_1)+\int^{\underline{k}^2_2}_{\underline{k}^2_{1{min}}}\frac{d\underline{k}^2_1}{\underline{k}^2_1}\int^1_{x_2}\frac{dx_1}{x_1}
{\cal
K}\left(\frac{\underline{k}^2_2}{\underline{k}_1^2},\frac{x_2}{x_1},\overline{\alpha}_s\right)
f(\underline{k}_1^2,x_1), \eqno(3.2) $$ where the factorization
scales are integrated, thus the observed cross section
$d\sigma(probe^*p\rightarrow k'X)$ is independent of the
factorization scheme.

    In the case of the evolution along the transverse momentum,
we differentiate Eq. (3.2) with respect to $\underline{k}^2_2$ and
get

$$\frac{\partial f(\underline{k}^2_2,x_2)}{\partial \underline{k}^2_2}$$
$$=\left .\int^1_{x_2}\frac{dx_1}{x_1}\frac{1}{\underline{k}^2_1}
{\cal
K}\left(\frac{\underline{k}_2^2}{\underline{k}_1^2},\frac{x_2}{x_1},\overline{\alpha}_s\right)
f(\underline{k}^2_1,x_1) \right |
_{\underline{k}^2_1=\underline{k}^2_2}$$
$$+\int^{\underline{k}^2_2}_{\underline{k}^2_{1{min}}}
\frac{d\underline{k}_1^2}{\underline{k}_1^2}
\int^1_{x_2}\frac{dx_1}{x_1}\frac{\partial {\cal
K}\left(\frac{\underline{k}_2^2}{\underline{k}_1^2},\frac{x_2}{x_1},\overline{\alpha}_s\right)}{\partial
\underline{k}^2_2} f(\underline{k}^2_1,x_1). \eqno(3.3)$$
Unfortunately, this equation is unsolvable in a resummation form.
However, at the $LL(\underline{k}^2)A$, the evolution kernel $\cal
K$ in Eq. (3.3) only is the function of the longitudinal variables
and the second term in the right-hand side of Eq. (3.3) vanishes.
In this case, using Eq. (2.12) we have

$$\Delta G(Q^2,x_2)\equiv\int^{Q^2}_{\underline{k}^2_{2min}}\frac{d\underline{k}^2_2}{\underline{k}^2_2}
\Delta f(\underline{k}_2^2,x_2)$$
$$=\int^{Q^2}_{\underline{k}^2_{2min}}\frac{d\underline{k}^2_2}{\underline{k}^2_2}
\int^{\underline{k}^2_2}_{\underline{k}^2_{1min}}\frac{d\underline{k}^2_1}{\underline{k}^2_1}
\int_{x_2}^1\frac{dx_1}{x_1}{\cal
K}\left(\frac{x_2}{x_1},\overline{\alpha}_s\right)f(\underline{k}^2_1,x_1)$$
$$\equiv\int^{Q^2}_{\underline{k}^2_{2min}}\frac{d\underline{k}^2_2}{\underline{k}^2_2}
\int_{x_2}^1\frac{dx_1}{x_1}\frac{x_2}{x_1} {\cal
K}_{DGLAP}\left(\frac{x_2}{x_1},\overline{\alpha}_s\right)G(\underline{k}^2_2,x_1),
\eqno(3.4)$$ and

$$G(Q^2,x_2)=G(\underline{k}^2_2,x_1)+\Delta G(Q^2,x_2),\eqno(3.5)$$
which lead to the DGLAP equation for the gluon distribution

$$Q^2\frac{\partial g(Q^2,x_2)}{\partial Q^2} =\int_{x_2}^1\frac{dx_1}{x_1}
{\cal K}_{DGLAP}\left(\frac{x_2}{x_1},\overline{\alpha}_s\right)
g(Q^2,x_1). \eqno(3.6)$$ We could extract the evolution kernel
${\cal K}_{DGLAP}$ in the collinear factorization scheme [14] if
the transverse momenta of the initial partons are neglected.

    It is interesting that the DGLAP equation Eq. (3.6) also can be
derived from the renormalization group, where the gluonic
structure function is factorized as

$$F(Q^2,x_2)=\int_{x_2}^1\frac
{dx_1}{x_1}C\left(\frac{Q}{\mu},\frac{x_2}{x_1},\alpha_s(\mu^2)\right)g(\mu^2,x_1).
\eqno(3.7)$$ In this equation $\mu$ is the unintegrated
renormalization scale as well as the factorization scale, $C$ is
the perturbative coefficient function.  Using

$$\mu\frac{\partial F(Q^2,x_2)}{\partial \mu}=0, \eqno(3.8)$$ one
can get [15]

$$\int^1_{x_2}\frac{dx_1}{x_1}
\left[\frac{dC\left(\frac{Q}{\mu},
\frac{x_2}{x_1},\alpha_s(\mu)\right)}{d\ln\mu^2}g(\mu^2,x_1)+C\left(\frac{Q}{\mu},
\frac{x_2}{x_1},\alpha_s(\mu)\right) \frac{dg(\mu^2,x_1)}{d\ln
{\mu}^2}\right]=0,\eqno(3.9)$$ at the lowest order of
$\alpha_s(\mu^2)$, it becomes

$$\int^1_{x_2}\frac{dx_1}{x_1}
C^{(0)}\left(\frac{Q}{\mu}, \frac{x_2}{x_1},\alpha_s(\mu)\right)
\frac{dg(\mu^2,x_1)}{d\ln {\mu}^2}$$
$$=-\int^1_{x_2}\frac{dx_1}{x_1}\frac{dC^{(1)}\left(\frac{Q}{\mu}, \frac{x_2}{x_1},\alpha_s(\mu)\right)}{d\ln\mu^2}g(\mu^2,x_1). \eqno(3.10)$$  Using

$$C^{(0)}=\delta\left(1-\frac{x_2}{x_1}\right), \eqno (3.11)$$
and

$$C^{(1)}={\cal K}_{GDLAP}\left(\frac {x_2}{x_1}\right)\ln \frac{Q^2}{\mu^2},
\eqno(3.12)$$ one can get the same results as Eq. (3.6).

     One difference between the coefficient function $C^{(1)}$ in
Eq. (3.10) and the evolution kernel ${\cal K}_{DGLAP}$ in Eq.
(3.6) is that two propagators connected with the probe in the
coefficient function $C^{(1)}$ are off mass-shell, but not on
mass-shell like in Eq. (3.1). However, as we have proved that the
evolution kernels from two methods are really equivalent in the
infrared safe theory even at twist-4.  The reason is that the
contributions of the backward components in these two propagators
always are suppressed [5].

    Considering that the scale-invariant parton picture [13] is a
phenomenological interpretation of the renormalization group in
the derivation of the QCD evolution equation, the above mentioned
two methods are really equivalent.

    We now construct the evolution equations along the longitudinal variable.
As we shall show that the transverse position of the targeted
gluon is uncertain in this case, we need to change the integrated
phase space in Eq. (3.2), i.e.,

$$\int\frac{d\underline{k}^2_1}{\underline{k}^2_1}\int\frac{dx_1}{x_1}\rightarrow
\int\frac{d\underline{k}^2_2}{\underline{k}^2_2}\int\frac{dx_2}{x_2}
\frac{\underline{k}^2_2}{\underline{k}^2_1}\frac{x_2}{x_1}.
\eqno(3.13)$$ Thus, we have

$$f(\underline{k}^2_1,x_1)=f(\underline{k}^2_2,x_2)-\Delta f(\underline{k}^2_1,x_1)$$
$$=f(\underline{k}^2_2,x_2)-\int\frac{d\underline{k}^2_2}{\underline{k}^2_2}\int^{x_1}_{x_{min}}\frac{dx_2}{x_2}
\frac{\underline{k}^2_2}{\underline{k}^2_1}\frac{x_2}{x_1}{\cal
K}\left(\frac{\underline{k}^2_2}{\underline{k}_2^1},\frac{x_2}{x_1},\overline{\alpha}_s\right)
f(\underline{k}_2^2,x_2)$$
$$ \equiv f(\underline{k}^2_2,x_2)-\int\frac{d\underline{k}^2_2}{\underline{k}^2_2}\int^{x_1}_{x_{min}}\frac{dx_2}{x_2}
{\cal
K}'\left(\frac{\underline{k}^2_2}{\underline{k}_2^1},\frac{x_2}{x_1},\overline{\alpha}_s\right)
f(\underline{k}_2^2,x_2), \eqno(3.14)$$  where we omit the
integral limit for the transverse momentum since they are
non-ordering.  At the $LL(1/x)A$, $\cal{K}'$ is only the function
of the transverse momenta.  We differentiate Eq. (3.14) with
respect to $x_1$ and obtain

$$-x\frac{\partial f(\underline{k}^2_1,x)}{\partial x}$$
$$=\int\frac{d\underline{k}_2^2}{\underline{k}_2^2}
{\cal
K}'\left(\frac{\underline{k}_2^2}{\underline{k}_1^2},\overline{\alpha}_s\right)
f(\underline{k}^2_2,x)$$
$$\equiv\int\frac{d\underline{k}_2^2}{\underline{k}_2^2} {\cal
K}_{BFKL}\left(\frac{\underline{k}_2^2}{\underline{k}_1^2},\overline{\alpha}_s\right)
f(\underline{k}^2_2,x), \eqno(3.15) $$  This is the real part of
the BFKL equation. Where $`-'$ implies that the evolution along
the direction of decreasing $x$, i.e., $x\rightarrow x-dx$.

    Now let us calculate the evolution kernel.  We take the square of the total amplitude in Fig. 1b and
separate out the probe vertex using the $W-W$ approximation [11],
one can get the four TOPT-diagrams Fig. 6, where the evolution
kernels are constructed by following amplitudes:

$$A_{BFKL}=A_{BFKL1}+A_{BFKL2},$$

$$A_{BFKL1}=\sqrt{\frac{E_{k}}{E_{p_a}}}\frac{1}{2E_{k}}
\frac{1}{E_{k}+E_{l_a}-E_{p_a}}M_{b1},$$ and

$$A_{BFKL2}=\sqrt{\frac{E_{k}}{E_{p_b}}}
\frac{1}{2E_{k}}\frac{1}{E_{k}+E_{l_b}-E_{p_b}}M_{b2}.\eqno(3.16)$$

    The momenta of the partons are parameterized as

$$p_a=(x_1P+\frac{(\underline{k}+\underline{l}_a)^2}{2x_1P}
,\underline{k}+\underline{l}_a,x_1P), \eqno(3.17)$$

$$k=(x_2P+\frac{\underline{k}^2}{2x_2P},\underline{k},x_2P),\eqno(3.18)$$

$$l_a=((x_1-x_2)P+\frac{\underline{l}_a^2}{2(x_1-x_2)P},\underline{l}_a,
(x_1-x_2)P), \eqno(3.19)$$

$$p_b=(x_1P+\frac{(\underline{k}
+\underline{l}_b)^2} {2x_1P} ,\underline{k}+\underline{l}_b,
x_1P), \eqno(3.20)$$ and

$$l_b=((x_1-x_2)P+\frac{\underline{l}_b^2}
{2(x_1-x_2)P}, \underline{l}_b, (x_1-x_2)P). \eqno(3.21)$$

    One of the matrix is

$$M_{BFKL1}=igf^{abc}[g_{\alpha\beta}(p_a+k)_{\gamma}+
g_{\beta\gamma}(-k+l_a)_{\alpha}+g_{\gamma\alpha}(-l_a-p_a)_{\beta}]
\epsilon_{\alpha}(p_a)\epsilon_{\beta}(k) \epsilon_{\gamma}(l_a),
\eqno(3.22)$$ where the polarization vectors are

$$\epsilon(p_a)=(0,\underline{\epsilon},
-\frac{\underline{\epsilon}\cdot(\underline{k}+\underline{l}_a)}{x_1P}),
\eqno(3.23)$$ and

$$\epsilon(k)=(0,\underline{\epsilon},
-\frac{\underline{\epsilon}\cdot\underline{k}}{x_2P}),\eqno(3.24)$$

$$\epsilon(l_a)=(0,\underline{\epsilon},
-\frac{\underline{\epsilon}\cdot\underline{l}_a}{(x_1-x_2)P}).
\eqno(3.25)$$

   Taking the leading logarithmic $(1/x)$ approximation, i.e.,
assuming that $x_2\ll x_1$, one can get the total amplitude

$$A_{BFKL}(\underline{k}_{up},\underline{k}_{down},x_1,x_2)=igf^{abc}2\sqrt{\frac{x_1}{x_2}}
[\frac{\underline{\epsilon}\cdot\underline{k}_{up}}{\underline{k}_{up}^2}
+\frac{\underline{\epsilon}\cdot\underline{k}_{dwon}}{\underline{k}_{down}^2}],
\eqno(3.26)$$  where we use different indices $up$ and $down$ to
distinguish between the different paths of the probing position
along the time order on the impact space.

    The impact parameter in the evolution equations is first
introduced by A. Mueller in his color dipole approaches [8]. We
imitate his method but work in the scattering model. After Fourier
transforming with respect to the transverse momentum transfer, we
obtain the representation of the amplitude in the impact parameter
space as follows

$$A_{BFKL}(\underline{x}_{a0},\underline{x}_{0b},x_1,x_2)=
\int{\frac{d^2\underline{k}_{up}d^2\underline{k}_{down}}{(2\pi)^4}
A_{BFKL}(\underline{k}_{up},\underline{k}_{down}},x_1,x_2)e^{i\underline{k}_{up}\underline{x}_{a0}
+i\underline{k}_{down}\underline{x}_{0b}}$$
$$=igf^{abc}2\sqrt{\frac{x_1}{x_2}}[\frac{\underline{x}_{a0}}{\underline{x}_{a0}^2}-
\frac{\underline{x}_{b0}}{\underline{x}_{b0}^2}]\cdot\underline{\epsilon}.
\eqno(3.27)$$ Using this amplitude we construct the evolution
kernel in the BFKL equation

$${\cal K}_{BFKL}\frac{d\underline{x}_{a0}^2}{\underline{x}_{a0}^2}\frac{dx_2}{x_2}=\sum_{pol}
A_{BFKL}A_{BFKL}^{\ast}\frac{d^3l_a}{(2\pi^3)2E_{l_a}}$$
$$=<3>_{color}\frac{\overline{\alpha}_{s}}{\pi^2}
\frac{\underline{x}_{ab}^2}{\underline{x}_{a0}^2\underline{x}_{b0}^2}d^2\underline{x}_{0}\frac{dx_2}{x_2},
\eqno(3.28)$$ where the position $\underline{x}_0$ in Eq. (3.28)
is running along the transverse coordinator due to uncertainty
relation although $\underline{k}_{up}$ identifies with
$\underline{k}_{down}$ in the forward deep inelastic scattering.
Therefore, the measurement of the unintegrated parton
distributions in the BFKL dynamics is non-local and we use the
dark box to indicate the above mentioned measurement. On the other
hand, we can not regard the fraction $x$ of the longitudinal
momentum of the parton in the BFKL equation as the Bjorken
variable since the transverse scale $\underline {k}^2$ of this
parton is irrelevant to the probe scale $Q^2$.

    Using Eqs. (3.15) and (3.28), we get

$$-x\frac {\partial f(\underline{x}_{ab}^2,x)}{\partial x}$$
$$=\frac{3\overline{\alpha}_{s}}{\pi^2}\int d^2 \underline{x}_0
\frac{\underline{x}_{ab}^2}{\underline{x}_{a0}^2\underline{x}_{b0}^2}
f(\underline{x}_0^2,x)$$
$$=\frac{3\overline{\alpha}_{s}}{2\pi^2}\int d^2 \underline{x}_0
\frac{\underline{x}_{ab}^2}{\underline{x}_{a0}^2\underline{x}_{b0}^2}
[f(\underline{x}_{a0}^2,x)+f(\underline{x}_{0b}^2,x)].\eqno(3.29)$$
Note that the distribution $f(\underline{x}_0^2,x)$ of a gluon at
$\underline{x}_0$ inside a cold spot, which is scaled by a fixed
$\underline{x}_{ab}$ can be represented by either
$f(\underline{x}_{a0}^2,x)$ (if taking the coordinate relative to
$\underline{x}_{a}$), or $f(\underline{x}_{0b}^2,x)$ (if taking
the coordinate relative to $\underline{x}_{b}$). The right-hand
side of Eq. (3.29) is illustrated in Fig. 7b, where the
un-connecting gluons in Fig. 7 means all possible connections as
shown in Fig. 6 and in the last diagram the probe is included.

    The equation (3.29) is the real part of the BFKL
equation in the impact space. A complete BFKL equation should
include the contributions from the virtual processes, which are
necessary for the infrared safety of the evolution dynamics. Using
the TOPT-cutting rule (see Appendix), one can prove that the
virtual diagrams in Fig 8 (where we omit the conjugation diagrams)
contribute the same evolution kernel as the real kernel but differ
by a factor $-1/2$. In the mean time, the probed distribution
$f(\underline{x}_{a0}^2,x)$ is replaced by
$f(\underline{x}_{ab}^2,x)$ because of the change of the cut line.
Thus, we can directly write the complete equation as

$$-x\frac {\partial f(\underline{x}_{ab}^2,x)}{\partial x}$$
$$=\frac{3\overline{\alpha}_{s}}{2\pi^2}\int d^2 \underline{x}_0
\frac{\underline{x}_{ab}^2}{\underline{x}_{a0}^2\underline{x}_{b0}^2}\left
[f(\underline{x}_{a0}^2,x)+f(\underline{x}_{b0}^2,x)-
f(\underline{x}_{ab}^2,x)\right]. \eqno(3.30)$$ This form of the
BFKL equation was first derived by Mueller in the color dipole
model [8].

    To make regularity of the evolution kernel explicit, we use
(for simplicity we replace $\underline{x}_{ab}^2\rightarrow \rho,
\underline{x}_{a0}^2\rightarrow \rho_1$ and
$\underline{x}_{b0}^2\rightarrow \rho_2)$

$$\frac{\rho}{\rho_1\rho_2}=\frac{1}{\rho_1}+\frac{1}{\rho_2}+\frac{\rho-\rho_1-\rho_2}{\rho_1\rho_2}.\eqno(3.31)$$
The evolution equation becomes

$$x\frac {\partial f(\rho,x)}{\partial x}$$
$$=\frac{3\overline{\alpha}_{s}}{\pi}\int_0^{\infty}\frac{d\rho_1}{\vert\rho_1-\rho\vert}
\left[\frac{\rho}{\rho_1}f(\rho_1,x)-f(\rho,x)\right]+\frac{3\overline{\alpha}_{s}}{\pi}\int_{\rho}^{\infty}
\frac{d\rho_1}{\rho_1}f(\rho,x). \eqno(3.32)$$ The first integral
of Eq. (3.32) is regular both for $\rho_1\rightarrow 0$ and
$\rho_1\rightarrow \rho$.  Defining $\rho_1=u\rho$ when
$\rho_1<\rho$ and $\rho_1=\rho/u$ when $\rho_1>\rho$, one can
write Eq. (3.32) as

$$-x\frac {\partial f(\rho,x)}{\partial x}$$
$$=\frac{3\overline{\alpha}_{s}}{\pi}\int_0^1\frac{du}{1-u}[f(u\rho,x)/u+f(\rho/u,x)-2f(\rho,x)].
\eqno(3.33)$$  The solution at small $x$ is power behaved

$$f(\underline{x}_{ab}^2,x)\sim \sqrt{\underline{x}^2_{ab}}x^{-\lambda}, \eqno(3.34)$$
where

$$\lambda=4\overline{\alpha}_s\ln2. \eqno(3.35)$$

     Now let us discuss the relation of the BFKL dynamics with
the DGLAP evolution equation.  The correlations among the initial
gluons disappear in the dilute parton system. In this case the
contributions of the interference diagrams Figs. 6c and 6d are
negligible. Thus, Eq. (3.28) reduces to sum two independent
splitting functions, each of which is the splitting function in
the DGLAP equation in the impact space at the small $x$ limit,

$${\cal K}_{DGLAP}\frac{d\underline{x}_{a0}^2}{\underline{x}_{a0}^2}\frac{dx_1}{x_1}=\sum_{pol}|A_{DGLAP}(\underline{x}_{a0},x_1,x_2)|^2
\frac{d^3l_a}{(2\pi^3)2E_{l_a}}$$
$$=\frac{6\overline{\alpha}_s}{2\pi}
\frac{dx_1}{x_2}\frac{d\underline{x}_{a0}^2}{\underline{x}_{a0}^2}.\eqno(3.36)$$
Because the interference amplitudes disappear in the DGLAP
dynamics, the transverse position of the probing parton is
irrelevant to the measurement, a simple probe (dashed line in Fig.
7) can be used. On the other hand, a single transverse scale
$\underline {x}^2_{a0}$ in the DGLAP dynamics allows us to take
$Q^2=\underline{k}^2$. Thus we can define the Bjorken variable
$x_B$ and set $x_2=x_B$ in Eq. (3.36). Returning to the momentum
space and using Eq. (3.6), one can get the real part of the DGLAP
equation at the double leading logarithmic (i.e., $\ln Q^2$ and
$\ln(1/x)$) approximation ($DLLA$)

$$\frac{\partial g(Q^2,x_B)}{\partial \ln Q^2}=
\int ^1_{x_B}\frac{dx_1}{x_1}{\cal
K}_{DGLAP}(z,\alpha_s)g(Q^2,x_1)-\frac{1}{2}g(Q^2,x_B)\int
^1_{0}dz{\cal K}_{DGLAP}(z,\alpha_s)$$
$$(small~x~limit)\longrightarrow\frac{\overline{\alpha}_s}{2\pi}\int
^1_{x_B}dx_1\frac{6}{x_B}g(Q^2,x_1),  \eqno(3.37)$$  where the
contributions of the virtual diagrams disappear at small $x$ since
they company with the singularities at $x\rightarrow 1$ (see
appendix A).

    The above derivations also reveal a relation between two evolution equations
from a new view point: the evolution kernel of the DGLAP equation
is the un-interferant part of the BFKL kernel.  Or reversely, the
BFKL equation can be regarded as the corrections from the initial
parton correlations to the DGLAP equation on the twist-2 level.

\newpage
\begin{center}
\section{Evolution equations at twist-4}
\end{center}

        We consider the evolution kernel based on Fig. 1d and call it as the modified BFKL equation.
Note that two pairs of initial gluons, which are hidden in the
correlation function, for example in Fig. 9a, should be indicated
as Fig. 9b. A set of cut diagrams based on the Fig. 1d are listed
in Fig. 10, where the probe vertices have been separated out using
the $W-W$ approximation. Similar to the derivation of Eq. (3.28),
we write the evolution kernel in this equation as

$${\cal K}_{MD-BFKL}=\sum_{pol}
A_{MD-BFKL}A_{MD-BFKL}^{\ast}\frac{d^3l_a}{(2\pi^3)2E_{l_a}}.
\eqno(4.1)$$ The amplitudes

$$A_{MD-BFKL}=A_{MD-BFKL1}+A_{MD-BFKL2}, \eqno(4.2)$$
where

$$A_{MD-BFKL1}=\sqrt{\frac{E_k}{E_{p_a}+E_{p_b}}}\frac{1}{2E_k}
\frac{1}{E_k+E_{l_a}-E_{p_a}-E_{p_b}}M_{MD-BFKL1},\eqno(4.3)$$ and

$$A_{MD-BFKL2}=\sqrt{\frac{E_k}{E_{p_c}+E_{p_d}}}
\frac{1}{2E_k} \frac{1}{E_k+E_{l_d}-E_{p_c} -E_{p_d}}M_{MD-BFKL2}.
\eqno(4.4)$$

For simplicity, in Fig. 10 we assume that all cold spots have a
same size

$$\underline{x}_{ab}=\underline{x}_{a'b'}=\underline{x}_{cd}=\underline{x}_{c'd'},$$
and

$$\underline{x}_{bc}=\underline{x}_{b'c'}. \eqno(4.5)$$

     The momenta of the partons, for example, are parameterized as

$$p_a=(x_1P+\frac{(\underline{l}_a-\underline{m})^2}{2x_1P}
,\underline{l}_a-\underline{m},x_1P), \eqno(4.6)$$

$$p_b=(x_1P+\frac{(\underline{k}+\underline{m})^2}{2x_1P}
,\underline{k}+\underline{m},x_1P), \eqno(4.7)$$

$$k_b=(x_2P+\frac{\underline{k}^2}{2x_2P},\underline{k},x_2P),\eqno(4.8)$$

$$l_a=((2x_1-x_2)P+\frac{\underline{l}_a^2}{2(2x_1-x_2)P},\underline{l},
(2x_1-x_2)P), \eqno(4.9)$$ and in the t-channel

$$m=p_b-k_b=((x_1-x_2)P+\frac{(\underline{k}+\underline{m})^2}{2x_1P}
-\frac{\underline{k}^2}{2x_2P},\underline{m}.(x_1-x_2)P).
\eqno(4.10)$$

    The matrix in Eq. (4.3) is

$$M_{MD-BFKL1}=igf^{abc}C^{\alpha\beta\gamma}\frac{-id^{\gamma\eta}_{\perp}}
{m^2}igf^{dce}C^{\rho\sigma\eta}\epsilon_{\alpha}(p_a)
\epsilon_{\rho}(p_b)\epsilon_{\beta}^{\ast}(l_a)
\epsilon^{\ast}_{\sigma}(k), \eqno(4.11)$$ where
$C^{\alpha\beta\gamma}C^{\rho\sigma\eta}$ are the triple gluon
vertices and polarization vectors are

$$\epsilon(p_a)=(0,\underline{\epsilon},
-\frac{\underline{\epsilon}\cdot(\underline{l}_a-\underline{m})}{x_1P}),\eqno(4.12)$$

$$\epsilon(p_b)=(0,\underline{\epsilon},
-\frac{\underline{\epsilon}\cdot(\underline{k}+\underline{m})}{x_1P}),\eqno(4.13)$$

$$\epsilon(k_b)=(0,\underline{\epsilon},
-\frac{\underline{\epsilon}\cdot\underline{k}}{x_2P}),
\eqno(4.14)$$ and

$$\epsilon(l_a)=(0,\underline{\epsilon},
-\frac{\underline{\epsilon}\cdot\underline{l}_a}{(2x_1-x_2)P}).
\eqno(4.15)$$ Thus, we have

$$A_{MD-BFKL}(\underline{k}_{up},\underline{k}_{down},\underline{m},
\underline{m}^{\prime},x_1,x_2)$$
$$=g^2f^{abc}f^{dce}\sqrt{\frac{x_1}{2x_2}}
[6\frac{\underline{\epsilon}\cdot\underline{k}_{up}
\underline{\epsilon}\cdot\underline{m}}{k_{up}^2m^2}
+6\frac{\underline{\epsilon}\cdot\underline{k}_{down}
\underline{\epsilon}\cdot\underline{m}^{\prime}}
{k_{down}^2m^{\prime 2}}]. \eqno(4.16)$$

    Taking the Fourier transformation

$$A_{MD-BFKL}(\underline{x}_{b0},\underline{x}_{0c},\underline{x}_{ab},
\underline{x}_{cd},x_1,x_2)$$
$$=\int{\frac{d^2\underline{k}_{up}d^2\underline{k}_{down}
d^2\underline{m}d^2\underline{m}^{\prime}}{(2\pi)^8}
A_{MD-BFKL}(\underline{k}_{up},\underline{k}_{down},\underline{m},
\underline{m}^{\prime},x_1,x_2)e^{i\underline{k}_{up}\underline{x}_{b0}
+i\underline{k}_{down}\underline{x}_{0c}+i\underline{m}
\underline{x}_{ab}
+i\underline{m}^{\prime}\underline{x}_{c^{\prime}d}}},
\eqno(4.17)$$ we get

$$A_{MD-BFKL}(\underline{x}_{b0},\underline{x}_{c0},\underline{x}_{ab}
,x_1,x_2)=6g^2f^{abc}f^{dce}\sqrt{\frac{x_1}{2x_2}}
[\frac{\underline{x}_{b0}\underline{x}_{ab}}{\underline{x}_{b0}^2\underline{x}_{ab}^2}
-\frac{\underline{x}_{c0}\underline{x}_{ab}}{\underline{x}_{c0}^2\underline{x}_{ab}^2}]
\underline{\epsilon}\cdot\underline{\epsilon}. \eqno(4.18)$$

Summing all channels, we derived the evolution kernel
corresponding to Fig. 10 as

$${\cal K}^{MD-BFKL}\frac{d\underline{x}_0^2}{\underline{x}_{a0}^2}\frac{d\underline{x}_{ab}^2}{\underline{x}_{ab}^2}\frac{dx_2}{x_2}$$
$$=\sum_{pol}
A_{MD-BFKL}A_{MD-BFKL}^{\ast}\left[\frac{1}{16\pi^3}\frac{dx_2}{x_2}d^2\underline{x}_0d^2\underline{x}_{ab}\right]$$
$$=2{\langle\frac{9}{32}\rangle}_{color}\frac{36\overline{\alpha}^2_s}{\pi}
\frac{d^2\underline{x}_0\underline{x}_{bc}^2}{\underline{x}_{b0}^2\underline{x}_{c0}^2}\frac{
d^2\underline{x}_{ab}}{\underline{x}_{ab}^2}\frac{dx_2}{x_2}.
\eqno(4.19)$$

    In the case of
$\vert\underline{x}_{bc}\vert\gg\vert\underline{x}_{ab}\vert$, one
can neglect the correlations between two cold spots with the sizes
$\underline{x}_{ab}$ and $\underline{x}_{cd}$. In consequence, the
contributions of the interferant terms (Figs. 10c and 10d)
disappear and Fig. 1d backs to Fig. 1c. Thus, Eq. (4.19) reduces
to the real part of the modified DGLAP-kernel

$${\cal K}^{MD-DGLAP}\frac{d\underline{x}_{b0}^2}{\underline{x}_{b0}^2}\frac{dx_1}{x_1}=2{\langle\frac{9}{32}\rangle}_{color}\frac{36\overline{\alpha}^2_s}{\pi}
\frac{d^2\underline{x}_{b0}}{\underline{x}_{b0}^4}\frac{dx_1}{x_2}.
\eqno(4.20)$$

      Similar to Eq. (3.5) we derive the modified DGLAP equation using

$$G(Q^2_2,x_2)=G(Q^2_1,x_1)+\Delta G(Q^2_2,x_2)$$
$$=G(Q^2_1,x_1)+\int^{Q^2_2}_{Q^2_{1min}}\frac{dQ^2_1}{Q^4_1}\int_{x_2}^{1/2}\frac{dx_1}{x_1}\frac{x_2}{x_1} {\cal
K}_{MD-DGLAP}\left(\frac{x_2}{x_1},\overline{\alpha}_s\right)
G^{(2)}(Q_1^2,x_1), \eqno(4.21)$$ where a power suppressed factor
$1/Q^2_1$ has been extracted from the evolution kernel. The
4-gluon correlation function $G^{(2)}$ is a generalization of the
gluon distribution beyond the leading twist. It is usually
modelled as the square of the gluon distribution [3,5]. For
example, $G^{(2)}= C_1G^2$ and $C_1=1/(\pi R^2_N)$. The evolution
constructed by ${\cal K}_{MD-DGLAP}$ is illustrated by Fig. 7c.

    The complete modified DGLAP equation combining DGLAP
dynamics at small $x$ and in the momentum space was written as [5]

$$\frac{\partial G(Q^2,x_B)}{\partial\ln Q^2}$$
$$=\frac{6\overline{\alpha}_s}{2\pi}\int^1_{x_B}
\frac{dx_1}{x_1} G(Q^2,x_1)
+\frac{81}{4}\frac{\overline{\alpha}_s^2}{\pi Q^2R^2_N}
\int_{x_B/2}^{1/2}\frac{dx_1}{x_1}G^{(2)}(Q^2,x_1)$$
$$-\frac{81}{2}\frac{\overline{\alpha}_s^2}{\pi Q^2R^2_N}
\int_{x_B}^{1/2}\frac{dx_1}{x_1}G^2(Q^2,x_1), \eqno(4.22)$$ where
the right-hand second term of (4.22) is positive antishadowing and
comes from the contributions from Fig. 11a, while the third term
is negative shadowing and arises from the contributions of the
$1\rightarrow 2$--$3\rightarrow 2$ cut diagrams (see Fig. 11b,
which is the same order as Fig. 11a but we haven't indicated it in
Fig. 1c). Note that the shadowing and antishadowing terms are
defined on different kinematics domains $[x,1/2]$ and $[x/2,
1/2]$, respectively.  Comparing with the GLR-MQ equation, there
are several features in the modified DGLAP equation: (i) The
momentum conservation of partons is restored in a complete
modified DGLAP equation; (ii) Because of the shadowing and
antishadowing effects in the modified DGLAP equation have
different kinematic regions, the net effect depends not only on
the local value of the gluon distribution at the observed point,
but also on the shape of the gluon distribution when the Bjorken
variable goes from $x$ to $x/2$. In consequence, the shadowing
effect in the evolution process will be obviously weakened by the
antishadowing effect if the distribution is steeper [5]; (iii)
Both the GLR-MQ and modified DGLAP equations assume that the
fusion only occurs among gluons with the same value of $x$.
However, there is a priori no reason to forbid the recombination
of two gluons with different values $x$. If the recombination of
gluons with different $x$ is included, we find that this
modification unreasonably enhances the shadowing effect in the
GLR-MQ equation, while it does not change the predictions of the
modified DGLAP equation. Therefore, a correct antishadowing
correction is important in gluon recombination [10]; (iv) Finally,
we emphasize that, as we have shown in [10], the existence in the
recombination process is a general result of the momentum
conservation and it is irrelevant to the concrete form of the
evolution dynamics.

    Now let us discuss the modified BFKL equation using Eq. (4.19), which
is the corrections of gluon recombination to the BFKL equation and
it is illustrated in Fig. 7d. Unfortunately, it is difficult to
write a simple form of the modified BFKL equation due to the
complicated evolution kernel (4.19). For the sake of
simplification, we take the following approximations (see Fig.
12): At the first step, we take the $W-W$ approximation to
factorize Fig. 12a to a four gluon correlation function (Fig. 12b)
and the general recombination function (Fig. 12c). Comparing Eq.
(3.36) with Eq. (3.28), one can find that the kernel ${\cal
K}_{DGLAP}$ is a part of the kernel ${\cal K}_{BFKL}$ and they
differ only by a phase space. Similarly, the modified DGLAP kernel
is evolving along the transverse momentum in the modified DGLAP
equation, but, as a part of the modified BFKL-kernel (Fig. 10a and
10b), they also participate the evolution along the longitudinal
momentum.  Therefore, at the next step, we only keep the
contributions of Fig. 10a in Fig. 12c, i.e., using the modified
DGLAP kernel to replace the kernel (4.19). According to Fig. 12,
we rewrite $\Delta G(Q^2_2,x_2)$ in Eq. (4.21) in the impact space
as

$$\Delta G(\underline{x}^2_{b0},x_2)=\int_{\underline{x}^2_{b0}}\frac{d\underline{x}^2_{ab}}{\underline{x}^4_{ab}}\int_{x_2orx_2/2}
^{1/2}\frac{dx_1}{x_1}\frac{x_2}{x_1} {\cal
K}_{MD-DGLAP}\left(\frac{x_2}{x_1},\overline{\alpha}_s\right)
f^{(2)}(\underline{x}_{ab}^2,x_1). \eqno(4.23)$$ Thus, we have

$$\Delta f(\underline{x}^2_{ab},x_1)=
\left.\underline{x}^2_{b0}\frac{\partial\Delta
G(\underline{x}^2_{b0},x_2)} {\partial \underline{x}^2_{b0}}
\right\vert_{\underline{x}^2_{ab}=\underline{x}_{b0}^2}$$
$$=\frac{1}{\underline{x}^2_{ab}}\int_{x_2orx_2/2}
^{1/2}\frac{dx_1}{x_1}\frac{x_2}{x_1} {\cal
K}_{MD-DGLAP}\left(\frac{x_2}{x_1},\overline{\alpha}_s\right)
f^{(2)}(\underline{x}_{ab}^2,x_1), \eqno(4.24)$$ which gives a
correction to the evolution of the unintegrated distribution along
small $x$ direction

$$-x\frac{\partial f(\underline{x}^2_{ab},x)}{\partial {x}^2_{ab}}
=\frac{81}{4}\frac{\overline{\alpha}_s^2}{\underline{x}_{ab}^2}
f^{(2)}(\underline{x}_{ab}^2,\frac{x}{2})
-\frac{81}{2}\frac{\overline{\alpha}_s^2}{x_{ab}^2}
f^{(2)}(\underline{x}_{ab}^2,x). \eqno(4.25)$$ Combining with the
BFKL equation, we obtain an approximate form of the modified BFKL
equation

$$-x\frac{\partial f(\underline{x}_{ab}^2,x)}{\partial x}$$
$$=\frac{\overline{\alpha}_{s}}{2\pi^2}\int d^2 \underline{x}_0
6\frac{\underline{x}_{ab}^2}{\underline{x}_{a0}^2\underline{x}_{b0}^2}f(\underline{x}_{a0}^2,x)-\frac{1}{2}\frac{\overline{\alpha}_{s}}{2\pi^2}
f(\underline{x}_{ab}^2,x)\int d^2 \underline{x}_06
\frac{\underline{x}_{ab}^2}{\underline{x}_{a0}^2\underline{x}_{b0}^2}$$
$$+\frac{81}{4}\frac{\overline{\alpha}_s^2}{\underline{x}_{ab}^2}
f^{(2)}(\underline{x}_{ab}^2,\frac{x}{2})
-\frac{81}{2}\frac{\overline{\alpha}_s^2}{x_{ab}^2}
f^{(2)}(\underline{x}_{ab}^2,x), \eqno(4.26)$$ where the nonlinear
evolution kernels are the derivative of the modified DGLAP kernels
with respect to $x$; in the meantime, we assume that $f^{(2)}=
C_2f^2$ and $C_2$ is a unknown correlation coefficient.

    Comparing with the modified DGLAP equation (4.22), the
antishadowing effect in the modified BFKL equation (4.26) is more
sensitive to the shape of the gluon distribution at small $x$. For
example, in the case of a typical BFKL solution Eq. (3.34), the
antishadowing effect may cancel, or even outwegh than the
shadowing effect in the modified BFKL equation (4.26). Such faster
increasing gluon distribution impels the multi-gluon recombination
(or correlation of multi-cold spots) to participate the evolution
and reach the saturation limit earlier.

    There is a difficulty in the numerical calculation of Eq. (4.26):
the value of $f( \underline{x}_{ab}^2,x/2)$ is unknown when the
equation evolves to small $x$. To avoid this problem, we take
following approximation at the beginning modified BFKL-renge, for
example, in the evolution from $x_1$ to $x_2=x_1-\Delta x$, the
increment of the unintegrated gluon distribution due to Eq. (4.26)
is taken as

$$\Delta f(\underline{x}_{ab}^2,x_1)=
BFKL[f(\underline{x}_{ab}^2,x_0)]+AS[f(\underline{x}_{ab}^2,\frac{x_0}{2})]
-S[f(\underline{x}_{ab}^2,x_0)]$$
$$\sim BFKL[f(\underline{x}_{ab}^2,x_0)]+AS[f_{BFKL}(\underline{x}_{ab}^2,\frac{x_0}{2})]
-S[f(\underline{x}_{ab}^2,x_0)], \eqno(4.27)$$ where $BFKL$, $AS$
and $S$ indicate the contributions from the BFKL kernel,
antishadwoing term and shadowing term in Eq. (4.26) and $f_{BFKL}$
implies the solution of the linear BFKL equation. A schematic
solution of Eq. (4.26) at this approximation is illustrated in
Fig. 13. Where we take a typical BFKL input distribution as

$$f(\underline {x}^2_{ab}, x_0)=\sqrt{\underline{x}_{ab}^2\underline{k}^2_s}
exp\left[-\frac{(log
(\underline{x}^2_{ab}\underline{k}^2_s)^2)}{5}\right],\eqno(4.28)$$
and $\underline{k}^2_s=1GeV^2$, $\underline {x}^2_{ab}=5
GeV^{-2}$. Comparing the coefficients of the nonlinear terms of
Eq. (4.26) with that of Eq. (4.22), we take
$C_2=1/(25\pi)GeV^{-2}.$

    The dashed-point curve in Fig. 13 is the imaginal saturation limit.
One can find that the antishadowing effect is un-negligible in
such abrupt distribution. For comparison, we draw a corresponding
BFKL-solution (dashed curve) and a solution of Eq. (4.26) without
the antishadowing corrections (point curve) in Fig. 13.

     The schematic kinematic ranges of four relating evolution equations
are shown in Fig. 14. Although both the BFKL equation and modified
DGLAP equation are regarded as the corrections of the correlations
among initial partons to the DGLAP dynamics, the BFKL equation
dominates the higher $Q^2$-range since the modified DGALP equation
is order of $\alpha^2_s$ and has a power suppression factor.
However, the modified DGLAP equation will replace the BFKL
equation below a moderate $Q^2$-scale because the $1/x$ factor in
${\cal K}_{MD-DGLAP}$ (see Eq. (4.20)). It is interesting that
because the gluon distribution becomes flatter in the modified
DGLAP-region in Fig. 14, the antishadowing contributions in Eq.
(4.26) can be neglected when the evolution enter into the modified
BFKL-region. Thus, we predict that the gluon distribution
asymptotically or abruptly approach to the saturation limit below
or above a moderate $Q^2$-scale.

\newpage
\begin{center}
\section{Discussions}
\end{center}

    At first, we point out that our nonlinear equation
(4.26) has a similar picture of the Balitsky-Kovchegov equation
[7], but they have different interpretations. The
Balitsky-Kovchegov equation is one of nonlinear evolution equation
incorporating with the screening corrections to the BFKL equation.
Besides, this equation is regarded as an approximation form of the
JIMWLK equation at the lowest order.  According to Ref. [8],
assuming that the scattering matrix $S(\underline{x}_{ab}, x)$,
which is the part of dashed box in Fig. 7b but use quark-antiquark
dipole to replace the gluonic dipole, obeys the BFKL equation

$$-x\frac {\partial S(\underline{x}_{ab}^2,x)}{\partial x}$$
$$=\frac{3\overline{\alpha}_{s}}{2\pi^2}\int d^2 \underline{x}_0
\frac{\underline{x}_{ab}^2}{\underline{x}_{a0}^2\underline{x}_{b0}^2}[S(\underline{x}_{a0}^2,x)+S(\underline{x}_{b0}^2,x)
-S(\underline{x}_{ab}^2,x)]. \eqno(5.1)$$ Then assuming that

(1) The gluon decay inside the dipole leads to the dipole
splitting at large $N_c$, i.e.,

$$S(\underline{x}_{a0}^2,x)+S(\underline{x}_{b0}^2,x)\rightarrow
S(\underline{x}_{a0}^2,x)S(\underline{x}_{b0}^2,x), \eqno(5.2)$$
in Eq. (5.1);

(2) Defining the scattering amplitude

$$S=1-T, \eqno(5.3)$$
one can simply obtain the Balitsky-Kovchegov equation

$$-x\frac {\partial T(\underline{x}_{ab}^2,x)}{\partial x}$$
$$=\frac{3\overline{\alpha}_{s}}{2\pi^2}\int d^2 \underline{x}_0
\frac{\underline{x}_{ab}^2}{\underline{x}_{a0}^2\underline{x}_{b0}^2}[T(\underline{x}_{a0}^2,x)+T(\underline{x}_{b0}^2,x)
-T(\underline{x}_{ab}^2,x)-T(\underline{x}_{a0}^2,x)T(\underline{x}_{b0}^2,x)],
\eqno(5.4)$$ where $T$ is relative to the gluon distribution [7].
Figure 15 is the one-step evolution containing the $4\rightarrow
2$ recombination amplitude.

    Comparing Fig. 15 with Fig. 7d, one can find that the Balitsky-Kovchegov equation
and the modified BFKL equation have a similar physical picture.
However, they are really different nonlinear evolution dynamics.
In fact, the replacements Eq. (5.2) and (5.3) result the negative
term $TT$ in Eq. (5.4). This term is graphically regarded as the
simultaneous scattering of two dipole. Such consideration of the
shadowing effect is relevant neither to the AGK- nor to
TOPT-cutting rules, but it is similar to a shadowing mechanism in
the multi-scattering process (for example, the Glauber-type
scattering ).

    Different from the recombination process, the antishadowing
effect is absent in the Glauber-type scattering since the shadowed
(or absorbed) momenta by target are un-measured [17]. Besides, the
linear parts and nonlinear part in the Balitsky-Kovchegov equation
share the same BFKL-kernel due to Eqs. (5.2) and (5.3). On the
other hand, the nonlinear evolution kernel in the modified BFKL
equation (4.26) is different from the linear BFKL-kernel since the
former including the recombination process. Thus, we present two
different shadowing mechanisms: the QCD gluon recombination
process and the absorptive effect in the multi-scattering process.

    We have shown a new approach to derive a set of consistent evolution equations in a
partonic framework.  We present the connections among the DGLAP
equation (3.37), BFKL equation (3.30), modified DGLAP equation
(4.22) and modified BFKL equation (4.26) in a clear physical
picture (Fig. 1). We also show that these evolution equations are
uniformly derived from the factorizability of DIS cross sections.
The four evolution equations mentioned above have similar
structures: The positive part in an evolution equation is
completely separated from the negative part. The positive part is
the contributions from the real diagrams, while the negative part
comes from the virtual diagrams in Eqs. (3.30) and (A-9), or from
the interference diagrams in Eqs. (4.22) and (4.26), respectively.

   The BFKL equation is traditionally understood as an improvement of
the DGLAP equation at small $x$, where the summation of the
leading logarithm of $x$ ($LL(1/x)$) terms is important. The
derivation of the BFKL equation in this work shows a new relation
with the modified DGLAP equation.  We re-draw the cut diagrams
Figs. 16a-16d in our derivation of the BFKL equation as the ladder
diagrams in Figs. 16e-16h. On the other hand, a typical diagram in
Regge approach of the BFKL equation is shown in Fig. 17, where the
vertical (dashed) lines are reggeized gluons, which can be
explained as the gluon ladders in perturbative QCD. Thus, Figs.
16e-16h give a possible partonic explanation of the Regge theory
for the BFKL equation.

       As is well known, the BFKL equation could also be derived in
the color dipole model [8]. Now let us compare the evolution
equations in our partonic model with that in the color dipole
model. We point out that these two approaches have different
partonic configurations. The dipole is constructed by two heavy
quarks, which represent a whole nucleon in the color dipole model,
while the cold spot in our approach is generally the multi-parton
cluster (as one of the constituents of the nucleon). In the
concrete, we compare Fig. 18a in our partonic scattering model
with the color dipole picture Figs. 18b or 18c, where the
transverse coordinators of the dipole are assumed to be freezed
during interaction. Note that the dashed lines in Fig. 18a are
integrated out as the final state in the inclusive DIS processes,
therefore, some of them are omitted in Fig. 1. On the other hand,
the splitting kernel in the color dipole model is constructed by
the gluon-quark vertex, while it is the three gluon vertex in our
partonic scattering approach. The latter dominates the small $x$
behavior since it  ($\sim 1/z_2$).

    It is interesting that a cut diagram Fig. 19a of the modified DGLAP
equation is equivalent to the ladder diagram Fig. 19b. Comparing
Fig. 19b with Figs. 17e-17h, we find that the BFKL and modified
DGLAP equations have the same (two ladders) structure in the
initial gluon configuration, but the probe partially couples with
two ladders in the BFKL equation at twist-2, while it fully
couples with two ladders though the gluon fusion mechanics in the
modified DGLAP equation at twist-4.

      We now discuss the relation of the BFKL dynamics
with the helicity configuration of the initial gluons.  As we have
pointed out in [5], the polarized form of two correlating initial
partons is relevant to the structure of the spin-independent
evolution kernel. For ease of representation, we draw all possible
helicity amplitudes in Fig. 20. Obviously, the processes
illustrated Figs. 20(a1)-20(a4) should be inhibited due to all
initial and final states in Fig. 20 having fixed helicities in the
$W-W$ approximation. Thus, the BFKL kernel (3.22) is a result,
where the contributions from Figs. 20(b1)-20(b4) being excluded,
i.e., we should assume that two gluons in a cold spot possess the
same helicity. On the other hand, the BFKL kernel reduces to the
DGLAP kernel if these two correlating gluons have the opposite
helicities, or the BFKL kernel becomes

$${\cal K}_{BFKL}\frac{d\underline{x}_0^2}{\underline{x}_{a0}^2}\frac{dx_2}{x_2}$$
$$=<3>_{color}\frac{\overline{\alpha}_{s}}{\pi^2}
\left[\frac{\underline{x}_{ab}^2}{\underline{x}_{a0}^2\underline{x}_{b0}^2}+\frac{2}{\underline{x}_{a0}^2}\right]
d^2\underline{x}_0\frac{dx_2}{x_2}, \eqno(5.5)$$ if two gluons in
a cold spot can take any helicities.

     In summary, we discussed the corrections of the initial
gluon correlations to the evolution dynamics in a unified
framework. We presented the following results: (1) a possible
connection among a set of evolution equations at small $x$; (2)
the antishadowing effect, which is a direct result of the momentum
conservation in elementary QCD process, is un-negligible in the
corrections of the gluon recombination to either the DGLAP or BFKL
equation; (3) the gluon distribution asymptotically (abruptly)
approaches to the saturation limit in lower (higher) $Q^2$ regions
is predicted; (4) the BFKL equation and modified DGLAP equation
are viewed as the corrections of the initial gluon correlations to
the evolution dynamics at twist-2 and twist-4, respectively; (5) a
partonic interpretation of Regge theory for the BFKL dynamics and
the relation of the BFKL dynamics with the helicity configuration
of the initial gluons are presented.

\vspace{0.3cm}

\noindent {\bf Acknowledgments}: This work was supported by
National Natural Science Foundations of China 10135060 and
10475028.

\newpage
\noindent {\bf Appendix}:

    For illustrating the applications of the TOPT-cutting rule,
which are proposed in [5], we write a complete derivation of the
DGLAP equation for gluon. The contributions of the real diagrams
(Fig. 21a) to the hard part of Eq. (3.1) are

$$H(probe^* l\rightarrow probe^*l)$$
$$=\frac{1}{2E_l}M_{l\rightarrow
kl'}\frac{1}{2E_k}\frac{1}{E_L-E_k-E_{l'}}\frac{1}{2E_k}\frac{1}{E_l-E_k-E_{l'}}M^*_{l\rightarrow
kl'}\frac{d^3l'}{(2\pi)^32E_{l'}}$$
$$\times \frac{1}{4E_{probe}}\vert M_{probe^*k\rightarrow
k'}\vert
^2(2\pi)^4\delta(p_{probe^*}+k-k')\frac{d^3k'}{(2\pi)^32E_{k'}}$$
$$={\cal K}_{DGLAP}(x_1,x_2,x_3, \alpha_s)dx_3\frac{dl^2_{\perp}}{l^2_{\perp}}\delta(x_2-x_B)dx_2C(probe^*k\rightarrow k'),
\eqno(A-1)$$ where the factor $1/(2E_l)$ is extracted from the
definition of the gluon distribution $g(Q^2,x_B)$, and

$${\cal K}_{DGLAP}(x_1,x_2,x_3, \alpha_s)dx_3\frac{dl^2_{\perp}}{l^2_{\perp}}$$
$$=\frac{E_k}{E_l}\vert M_{l\rightarrow
kl'}\vert^2\left[\frac{1}{E_l-E_k-E_{l'}}\right]^2\left[\frac{1}{2E_k}\right]^2\frac{d^3l'}{(2\pi)^32E_{l'}},\eqno(A-2)$$

$$C(probe^*k\rightarrow k')\delta(x_2-x_B)dx_2$$
$$=\frac{1}{8E_kE_{probe^*}}\vert M_{probe^*k\rightarrow
k'}\vert^2(2\pi)^4\delta(p_{probe^*}+k-k')\frac{d^3k'}{(2\pi)^32E_{k'}},
\eqno(A-3)$$ is the contributions from
$d\sigma(probe^*k\rightarrow k')$.  Thus,

$$\Delta g(Q^2,x_B)=\int
\frac{dl^2_{\perp}}{l^2_{\perp}}dx_1dx_2dx_3{\cal
K}_{DGLAP}(x_1,x_2,x_3,\alpha_s)g(l^2_{\perp},x_1)\delta(x_2-x_B)\delta(x_1-x_2-x_3),
\eqno(A-4)$$  where we inset $\delta(x_1-x_2-x_3) dx_1$.  We
obtain

$$\frac{\partial g(Q^2,x_B)}{\partial \ln Q^2}$$
$$=\int dx_1dx_2dx_3{\cal
K}_{DGLAP}(x_1,x_2,x_3,\alpha_s)g(Q^2,x_1)\delta(x_2-x_B)\delta(x_1-x_2-x_3)$$
$$=\int  dx_1dz{\cal K}_{DGLAP}(z,\alpha_s)g(Q^2,x_1)\delta(x_1z-x_B)$$
$$=\int \frac{dx_1}{x_1}{\cal K}_{DGLAP}(\frac{x_B}{x_1},\alpha_s)g(Q^2,x_1)\ . \eqno(A-5)$$

    On the other hand, the contributions from one of the virtual
diagrams, for say, Fig. (21b) are

$$H(probe^*l\rightarrow probe^*l)$$
$$=\frac{1}{2}\frac{1}{2E_l}M_{l\rightarrow
kl'}\frac{1}{2E_k}\frac{1}{2E_{l'}}\frac{1}{E_L-E_k-E_{l'}}M_{kl'\rightarrow
l}\frac{1}{2E_l}\frac{1}{E_k+E_{l'}-E_l}\frac{d^3l'}{(2\pi)^3}$$
$$\times \frac{1}{4E_{probe}}\vert M_{probe^*l\rightarrow
k'}\vert
^2(2\pi)^4\delta(p_{probe^*}+l-k')\frac{d^3k'}{(2\pi)^32E_{k'}}$$
$$=-\frac{1}{2}{\cal K}_{DGLAP}(x_1,x_2,x_3,\alpha_s)dx_3\frac{dl^2_{\perp}}{l^2_{\perp}}\delta(x_1-x_B)dx_1C(probe^*l\rightarrow k'),
\eqno(A-6)$$ where

$$C(probe^*l\rightarrow k')\delta(x_1-x_B)dx_1$$
$$=\frac{1}{8E_lE_{probe^*}}\vert M_{probe^*l\rightarrow
k'}\vert^2(2\pi)^4\delta(p_{probe^*}+l-k')\frac{d^3k'}{(2\pi)^32E_{k'}},
\eqno(A-7)$$ is the contributions from
$d\sigma(probe^*l\rightarrow k')$.  Therefore, we have

$$\frac{\partial g(Q^2,x_B)}{\partial \ln Q^2}$$
$$=-\frac{1}{2}\int dx_1dx_2dx_3{\cal
K}_{DGLAP}(x_1,x_2,x_3,\alpha_s)g(Q^2,x_1)\delta(x_1-x_B)\delta(x_1-x_2-x_3)$$
$$=-\frac{1}{2}g(Q^2,x_B)\int dz{\cal K}_{DGLAP}(z,\alpha_s)\ . \eqno(A-8)$$

    The complete evolution equation for the gluons is

$$\frac{\partial g(Q^2,x_B)}{\partial \ln Q^2}=\int
^1_{x_B}\frac{dx_1}{x_1}{\cal
K}_{DGLAP}(z,\alpha_s)g(Q^2,x_1)-\frac{1}{2}g(Q^2,x_B)\int
^1_{0}dz{\cal K}_{DGLAP}(z,\alpha_s), \eqno(A-9)$$ where

$${\cal
K}_{DGLAP}(z)=\frac{\alpha_s}{\pi}C_A\left[\frac{z}{1-z}+\frac{1-z}{z}+z(1-z)\right],\eqno(A-10)$$
Note that the factor $1/2$ in Eq. (A-8) is the considerations of
the symmetry under exchange of two internal gluons in the virtual
diagrams (see Ref. [5]). The expressions of the evolution equation
in the TOPT form, without the calculations of the matrixes, show
that the real and virtual diagrams contribute the same evolution
kernel but with the different factors. This simply form of the
equation consists with a more complicate derivation of the DGLAP
equation using the covariant perturbation theory in [19]. In fact,
from Eqs. (A-9) and (A-10) we have

$$\frac{\partial x_Bg(Q^2,x_B)}{\partial \ln Q^2}=\frac{3\alpha_s}{\pi}\int
^1_{x_B}dz\left[\frac{z}{1-z}+\frac{1-z}{z}+z(1-z)\right]
x_1g(Q^2,x_1)$$
$$-\frac{1}{2}\frac{3\alpha_s}{\pi}x_Bg(Q^2,x_B)\int
^1_{0}dz\left[\frac{z}{1-z}+\frac{1-z}{z}+z(1-z)\right],
\eqno(A-11)$$ Note that $\int^1_0dzz/(1-z)=\int^1_0dz(1-z)/z$, we
obtain

$$\frac{\partial x_Bg(Q^2,x_B)}{\partial \ln Q^2}=\frac{3\alpha_s}{\pi}\int
^1_{x_B}dz\left[\frac{zx_1g(Q^2,x_1)-x_Bg(Q^2,x_B)}{1-z}+\frac{(1-z)(1+z^2)}{z}x_1g(Q^2,x_1)\right]$$
$$-\frac{3\alpha_s}{\pi}x_Bg(Q^2,x_B)\int
^1_{0}dz\left[\frac{z}{1-z}+\frac{1}{2}z(1-z)\right]+\frac{3\alpha_s}{\pi}x_Bg(Q^2,x_B)\int
^1_{x_B}dz\frac{1}{1-z}$$
$$=\frac{3\alpha_s}{\pi}\int
^1_{x_B}dz\left[\frac{x_1g(Q^2,x_1)z-x_Bg(Q^2,x_B)}{1-z}+\frac{(1-z)(1+z^2)}{z}x_1g(Q^2,x_1)\right]$$
$$+\frac{\alpha_s}{\pi}\left[\frac{11}{4}+3\ln
(1-x)\right]x_Bg(Q^2,x_B). \eqno(A-12)$$ This equation is
equivalent to the following traditional for of the DGLAP equation

$$\frac{\partial g(Q^2,x_B)}{\partial \ln Q^2}=\int
^1_{x_B}\frac{dx_1}{x_1}{\cal
K}_{DGLAP}^{tradoctional}(z,\alpha_s)g(Q^2,x_1)), \eqno(A-13)$$
where

$${\cal
K}_{DGLAP}^{tradictional}(z)=\frac{\alpha_s}{\pi}C_A\left[\left(\frac{z}{1-z}\right)_++\frac{1-z}{z}+z(1-z)+\delta(1-z)\frac{11}{12}\right].\eqno(A-14)$$
In fact, using

$$\int^1_0dz\frac{f(z)}{(1-z)_+}\equiv\int^1_0dz\frac{f(z)-f(1)}{1-z},
\eqno(A-15)$$ Equation (A-14) becomes

$$\frac{\partial x_Bg(Q^2,x_B)}{\partial \ln Q^2}=\frac{3\alpha_s}{\pi}\int
^1_{x_B}dz\left[\frac{z}{1-z} +\frac{1-z}{z}+z(1-z)\right]$$
$$+\frac{11}{4\pi}\alpha_sx_Bg(Q^2,x_B)
-\frac{3\alpha_s}{\pi}\delta(1-z)\int^1_0dz\frac{1}{1-z}$$
$$=\frac{3\alpha_s}{\pi}\int
^1_{x_B}dz\left[\frac{x_1g(Q^2,x_1)z-x_Bg(Q^2,x_B)}{1-z}+\frac{(1-z)(1+z^2)}{z}x_1g(Q^2,x_1)\right]$$
$$+\frac{\alpha_s}{\pi}\left[\frac{11}{4}+3\ln
(1-x)\right]x_Bg(Q^2,x_B). \eqno(A-16)$$

    Both Eqs. (A-12) and (A-16) at small $x$ predict Eq. (3.37).
One can find that the infrared divergence of the real diagram at
the end point $(x_B\rightarrow 1)$ is exactly cancelled by the the
contributions of the virtual diagram. Therefore, the leading terms
of the virtual diagram in Eq. (3.37) disappear at the small $x$
region. It is different from Eq. (3.37), the contributions of the
virtual diagram exist in Eq. (3.30), which regularize the
singularities of the real diagram on the transverse space. The
above results also can be applied in the BFKL dynamics. A
difference in this case is that the absence of the conditions
$\delta(x_1-x_B)$ and $\delta(x_2-x_B)$ in Eqs. (A-1) and (A-6),
respectively.

\newpage

Figure Captions

Fig. 1  The corrections of the initial gluons to the basic
amplitude of the DGLAP equation and they lead to (b) BFKL-, (c)
modified DGLAP- and (d) modified BFKL equations, respectively.

Fig. 2  Application of the $W-W$ approximation. The dark circles
denote perturbative QCD interaction with the correlation of the
initial partons.

Fig. 3  A sub-vertex in a complex Feynman diagram.

Fig. 4  The parton distributions on the impact space of a nucleon:
(a) the dilute parton system, (b) some of partons form the cold
spots (dashed circles) and (c) the saturation, where the long- and
short-distance correlations among partons coexist in a nucleon.

Fig. 5  The QCD correlations between two initial gluons.

Fig. 6  The TOPT-diagrams consisted by the elemental amplitudes in
Fig. 1b on the impact space. These diagrams lead to the BFKL
equation. The dashed lines are the time ordered lines in the TOPT
and the dark box represent a non-local measurement of the
unintegrated distribution.

Fig. 7  The illustration of the right-hand side (real part) of (a)
DGLAP equation, (b) BFKL equation, (c) modified DGLAP equation and
(d) modified BFKL equation, where un-connecting gluons imply all
possible connections inside the cold spot like in Fig. 6. The last
sub-diagrams are added probe, which contribute $\delta(x-2-x_B)$
or Eq. (2.10) in (a) and (c), or in (b) and (d), respectively.

Fig. 8  The virtual diagrams corresponding to Fig. 6.

Fig. 9 A cutting diagram originating from Fig. 1d.

Fig. 10  The TOPT-diagrams consisted by the elemental amplitudes
in Fig. 1d in the impact space. These diagrams lead to the
modified BFKL equation.

Fig. 11  Two cutting diagrams contributing to the modified DGLAP
equation.

Fig. 12  The factorization of Fig. 10, which leads to an
approximate form of the modified BFKL equation.

Fig. 13  The unintegrated gluon distribution
$f(\underline{x}^2_{ab},x)$ at $\underline{x}^2_{ab}=5 GeV^{-2}$
as a solution of the modified BFKL equation (4.26) taking the
approximation (4.27) (solid curve), a corresponding solution of
the BFKL equation (dashed curve) and a solution of Eq. (4.26) but
without antishadowing (point curve). The broken-point line is an
imaginal saturation limit.

Fig. 14  A schematic kinematic region of four evolution equations
in this work.

Fig. 15  One-step evolution including the $4\rightarrow 2$
recombination amplitude, which is regarded as a graphic
description of the nonlinear terms in the Balitsky-Kovchegov
equation.

Fig. 16  Schematic cut diagrams for the BFKL equation (a-d) and
corresponding ladder diagrams (e-h).

Fig. 17  A typical ladder diagram in the traditional BFKL theory,
where the vertical (dashed) lines are reggeized gluons.

Fig. 18  A comparison of our partonic picture with the color
dipole approaches.

Fig. 19  (a) A cut diagrams of the modified DGLAP equation and (b)
corresponding ladder diagram.

Fig. 20 The helicity amplitudes contributing to the
spin-independent BFKL kernel, where the processes in (a1)-(a4)
should be inhibited at the $W-W$ approximation.

Fig. 21 Complete TOPT diagrams containing the probe for the DGLAP
equation: (a) real diagram and (b) virtual diagrams.

\end{document}